\documentclass[11pt]{article}

\usepackage{indentfirst}
\usepackage{amsfonts}
\usepackage{amsmath}
\usepackage{amsthm}
\usepackage{amssymb}
\usepackage{lineno}

\usepackage{verbatim}
\usepackage[makeroom]{cancel}
\usepackage{rotating}
\usepackage{float}
\usepackage{cite}
\usepackage{mathtools}
\usepackage{amssymb}
\usepackage{mathrsfs}
\usepackage{epsfig}
\usepackage{graphicx}
\usepackage{indentfirst}
\usepackage{framed}
\usepackage{xcolor}
\usepackage[backref=page]{hyperref}
\usepackage[capitalize]{cleveref}
\usepackage{xparse}

\usepackage{enumitem}
\usepackage{bm}
\usepackage[normalem]{ulem}
\usepackage{revsymb}

\usepackage{upgreek}

\usepackage[labelfont=bf]{caption}
\usepackage{soul}
\newcommand{\emptyword}{\epsilon}

\newcommand{\ket}[1]{|#1\rangle}

\newcommand{\braket}[2]{\langle#1|#2\rangle}

\newcommand{\ketbra}[2]{|#1\rangle\! \langle #2|}

\newcommand{\Tr}{\operatorname{Tr}}
\newcommand{\tr}[1]{\Tr\left(#1\right)}

\def\01{\{0,1\}}

\newcommand{\uu}{{\bf u}}
\newcommand{\vv}{{\bf v}}

\newcommand{\PSD}{\mathcal{S}^+}

\newcommand{\B}{\mathcal{B}}
\renewcommand{\H}{\mathcal{H}}
\newcommand{\BH}{{\mathcal B(\mathcal H)}}

\newcommand{\supp}{\operatorname{supp}}
\newcommand{\W}{\mathcal{W}}
\newcommand{\V}{\mathcal{V}}

\newcommand{\sa}{\mathrm{sa}}

\usepackage[margin=0.7in]{geometry}

\definecolor{corlinks}{RGB}{200,0,0}
\definecolor{cormenu}{RGB}{200,0,0}
\definecolor{corurl}{RGB}{200,0,0}

\hypersetup{
colorlinks=true,
urlcolor=corlinks,
linkcolor=corlinks,
menucolor=cormenu,
citecolor=corlinks,
pdfborder= 0 0 0
}
\theoremstyle{plain}

\newtheorem{theorem}{Theorem}

\newtheorem{definition}[theorem]{Definition}

\newtheorem{proposition}[theorem]{Proposition}

\theoremstyle{remark}

\def\01{\{0,1\}}

\DeclareDocumentCommand{\dist}{o}{%
  \IfNoValueTF{#1}{d}{d_{\mathrm{#1}}}%
}

\NewDocumentCommand{\Prob}{e{_} m}{%
  \IfNoValueTF{#1}{%
    \Pr \set*{#2}
  }{%
    \Pr_{#1} \set*{#2}
  }
}

\newcommand{\C}{\ensuremath{\mathcal{C}}}

\newcommand{\1}{\openone}

\begin{document}
\title{\vspace{-20pt}\huge \bf Quantum theory in finite dimension \protect\\ cannot explain every general process \protect\\ with finite memory}

\author{Marco Fanizza\thanks{{F\'{\i}sica Te\`{o}rica: Informaci\'{o} i Fen\`{o}mens Qu\`{a}ntics, Departament de F\'{i}sica, Universitat Aut\`{o}noma de Barcelona, ES-08193 Bellaterra (Barcelona), Spain}. \texttt{marco.fanizza@uab.cat}} 
\and Josep Lumbreras\thanks{{Centre for Quantum  Technologies,  National University of Singapore, Singapore}. \texttt{josep.lumbreras@u.nus.edu}} 
\and Andreas Winter\thanks{{ICREA \&{} F\'{\i}sica Te\`{o}rica: Informaci\'{o} i Fen\`{o}mens Qu\`{a}ntics, Departament de F\'{i}sica, Universitat Aut\`{o}noma de Barcelona, ES-08193 Bellaterra (Barcelona), Spain. AW is Hans Fischer Senior Fellow with the Institute for Advanced Study, Technische Universit\"at M\"unchen, Lichtenbergstra{\ss}e 2a, D-85748 Garching, Germany}. \texttt{andreas.winter@uab.cat}}}

\date{28 November 2022}
\maketitle
\vspace{-8mm}

\abstract
{Arguably, the largest class of stochastic processes generated by means of a finite memory consists of those that are sequences of observations produced by sequential measurements in a suitable generalized probabilistic theory (GPT). These are constructed from a finite-dimensional memory evolving under a set of possible linear maps, and with probabilities of outcomes determined by linear functions of the memory state. Examples of such models are given by classical hidden Markov processes, where the memory state is a probability distribution, and at each step it evolves according to a non-negative matrix, and hidden quantum Markov processes, where the memory state is a finite dimensional quantum state, and at each step it evolves according to a completely positive map. Here we show that the set of processes admitting a finite-dimensional explanation do not need to be explainable in terms of either classical probability or quantum mechanics. To wit, we exhibit families of processes that have a finite-dimensional explanation, defined manifestly by the dynamics of explicitly given GPT, but that do not admit a quantum, and therefore not even classical, explanation in finite dimension. Furthermore, we present a family of quantum processes on qubits and qutrits that do not admit a classical finite-dimensional realization, which includes examples introduced earlier by Fox, Rubin, Dharmadikari and Nadkarni as functions of infinite dimensional Markov chains, and lower bound the size of the memory of a classical model realizing a noisy version of the qubit processes.}

\medskip
\noindent 
\section{Introduction}
Modeling a hidden cause mechanism for the probability distribution of a time series of observations is a ubiquitous task, from fundamental science experiments to data analysis. Considering classical hidden dynamics gives rise to hidden Markov models (HMM)~\cite{Rabiner1989, Vidyasagar2014}, which have key applications in fields where time series arise~\cite{zucchini2009hidden}, among them speech recognition~\cite{jelinek1998statistical} and genomics~\cite{Burge1997}, where they are still an important part of the data analysis tools in these fields~\cite{Ernst2017}, but also new possible uses are emerging, such as in ecology~\cite{Glennie2022}. 
On the other hand, repeated measurements on a quantum system also define probabilities of sequences of outcomes with a hidden mechanism, in this case a quantum one. Landmark experiments can be modeled as such~\cite{Raimond2001}.  Infinite sequences of identical repeated measurements define the class of {hidden quantum Markov models (HQMM), a special case of C${}^*$-finitely correlated state when the state is classical, i.e. diagonal in a given product basis \cite{Fannes1992}}. {HQMMs not only can} serve as tools for the analysis of quantum experiments and for the modeling quantum technologies, but also as tools for data analysis application, implemented in a classical simulator or on actual controllable quantum systems (be it NISQ devices or universal quantum processors). 

Removing the restriction to classical or quantum dynamics, and keeping only on the linearity of the hidden dynamics and the nonnegativity of the function used to compute the probabilities of sequences, enlarges the class of possible models and ensuing processes to so-called quasi-realizations~\cite{Vidyasagar2014}. These generalized models are known under several different names in different communities, e.g. operator observable models (OOM)~\cite{Jaeger2000} or weighted finite automata~\cite{Balle2015}, or indeed (classical) finitely correlated states \cite{Fannes1992}. Considering this extended class simplifies greatly the inference of the hidden mechanism from the probabilities of the sequences, as a minimal description can be obtained by simple linear algebra, while this is not the case for a classical or quantum one. Moreover, from a physical point of view, this extended space of models can be seen as the class of models describing repeated measurement on a system in general probabilistic theories (GPTs)~\cite{Muller2021}, including alternatives or extensions of quantum theory. {The immediate question presenting itself} is whether there is a strict inclusion between the sets of HMM, {HQMM} and general models? For these sets and for any other subclass of models that can be conceived, this is an interesting question from a fundamental point of view, since one could say that the possibility of generating every stochastic process with finite memory is a desirable property of a general theory of nature, but it also has practical consequences for applications, since it can exhibit strengths or limitations of specific classes. Already in~\cite{fox1968,dharmadhikari1970} it was shown that there exist processes admitting general models which however are not representable classically by any HMM. In~\cite{monras2016} it was shown that there exist processes given by HQMM which however cannot be represented by classical HMM. Perhaps then quantum mechanics is sufficiently powerful to be able to realize any discrete process admitting a finite memory general model, by means of finite-dimensional quantum systems~\cite{monras2016}? 

The main contribution of the present paper is a negative answer to this question, via the explicit construction of processes admitting a general linear model, but for which the underlying possible GPT is so tightly constrained that we can exclude the possibility of a realization by HQMM by inspection. Our result also answers a question raised in \cite[Sec.~7.1]{Fannes1992}.
{The argument is geometric, as pioneered  in \cite{monras2016} (there for separating HMM and HQMM): our examples are such that the GPTs of their quasi-realizations have unique mutually dual convex cones of effects and states, respectively; in other words, there is only one possible operational probabilistic theory that can describe the observable statistics. As HQMM give rise to semi-definite representable (SDR cones, i.e. projections of sections of the positive-semidefinite cone of matrices), we can exclude a quantum realisation by forcing our cone to be not semi-algebraic.}
On the other hand, to better appreciate the power of HQMM and motivating the question of establishing a separation with general theories, we show that the non-classical examples in~\cite{fox1968,dharmadhikari1970} are representable by HQMMs, and thus are not sufficient to show the new separation. This is remarkable since these examples were naturally formulated as a functions of infinite-alphabet classical Markov models, showing that small quantum systems can be expressive enough to represent rich stochastic processes that are not inherently quantum, supporting the possibility that quantum systems can be useful for modeling real world data streams. On the other hand, by simplifying the original examples, we remark that already a class of binary sequential measurements on a qubit cannot be reproduced by a HMM. This fact was already noticed by~\cite{srinivasan2018learning} where a HQMM for the so-called probability clock of Jaeger~\cite{Jaeger2000,Zhao2010} was found, which itself is a simplified version of the older example in~\cite{fox1968,dharmadhikari1970}. 

Before going into a mathematically precise description of our framework and results, let us discuss further related work. The notion of quantum hidden Markov models seem to have appeared in~\cite{Wiesner2008}. In~\cite{Monras2010} a process was constructed which can be represented on a qubit but not on a binary classical space. Several papers analyzed how, for a quantum process representing a hidden Markov model, the entropy of the average stationary state can be less than in the classical case~\cite{Gu2012, Aghamohammadi2018, Elliott2020,Elliott2021a,Elliott2021b}, and how to construct a quantum representation of an HMM, or from the outcome probabilities~\cite{Liu2019,Ho2020}. In particular, an example of a class of classical processes which require infinite memory in a so-called \emph{unifilar} HMM, but can be implemented on a qubit, was shown in~\cite{Elliott2020}. A gap between the memory requirement of an $\epsilon$-machine to simulate sequential measurements in contextuality experiments was also observed~\cite{Cabello2018}.
Note however that it is well-known that there exist processes generated by a finite HMM, yet its  $\epsilon$-machine and any other unifilar HMM necessarily have infinite memory \cite{Crutchfield1994,MarzenCrutchfield205}.
The non-asymptotic behaviour of the sample mean of a HQMM has been studied in~\cite{hayashi2018} giving bounds for the tail probabilities and deriving a central-limit theorem type result. Algorithms to find a HQMM modelling a sequence of observations have been presented in~\cite{srinivasan2018learning,pmlr-v108-adhikary20a}. Note that HQMM can equivalently be obtained from locally measuring C${}^*$-finitely correlated states~\cite{Fannes1992}; this implies that our work also shows the existence of finitely correlated states which are not C${}^*$-finitely correlated, answering an open questions of~\cite{Fannes1992}, which received attention and but not a conclusive answer. For example,~\cite{Cuevas2013} shows that a similar separation exists for sequences of finite-size states in the non-translation invariant setting, while~\cite{Cuevas2016} shows that a separation exists for sequences of periodic finite-size states. Moreover, several works have investigated the use and advantages of tensor networks for probabilistic modeling, e.g.~\cite{NEURIPS2019_b86e8d03,Gao2022,pmlr-v130-miller21a,pmlr-v130-adhikary21a}. 

The cones used to show the separation are the power cone and the exponential cone~\cite{chares2009cones}, being the power cone more general since the exponential cone can be obtained as a limiting case of the power cone plus a linear transformation. They have no clear physical interpretation as general probabilistic theories (yet), but appear as models for several practical optimization problems, with applications to chemical process control~\cite{wall1986or}, circuit design~\cite{boyd2005digital}, or electric vehicle charging~\cite{chen2021exponential}, among many others. Both the power cone and the exponential cone have self-concordant barriers~\cite{chares2009cones,nesterov2006constructing,nesterov2012towards}  which make them suitable for conic optimization methods like interior point algorithms, and although they are non-symmetric cones the implementation of the algorithms is feasible~\cite{dahl2021primal}. The exponential cone also can be used to model relative entropy programs which includes geometric programming~\cite{boyd2007tutorial}  and second order conic programming~\cite{lobo1998applications}. Extensions to quantum relative entropy programs include tasks like quantum channel capacity approximation~\cite{chandrasekaran2017relative} or quantum state tomography~\cite{gonccalves2013quantum}.

The paper is organized as follows. In the results section we start by reviewing key properties of finite-dimensional linear models for stochastic processes, and of their classical and quantum realizations. Then we show that the processes in~\cite{fox1968,dharmadhikari1970} which do not admit a classical realization, do in fact admit a quantum realization. Moreover, we quantitatively evaluate the robustness of this statement by considering perturbation of the quantum realizations of these processes by depolarizing noise. We then present our main result: two families of processes with a three dimensional quasi-realization, which we show however not to admit any finite dimensional quantum realizations. Finally, in the discussion section, we present generalizations of the convex state spaces of the GPTs underlying the models, which also extend quantum theory.

\section{Results}
\subsection{Stationary stochastic processes and quasi-realizations}

We start by reviewing the formalism for general linear models with memory of stochastic processes, or quasi-realizations~\cite{Vidyasagar2014}. Let $\mathbb{M}$ be an alphabet with $|\mathbb{M}|=m$ symbols and let $\mathbb{M}^{\ell}$ be the set of words of length $\ell$. This includes $\ell=0$, in which case $\mathbb{M}^0$ consists only of one word $\epsilon$. By $\mathbb{M}^* = \bigcup_{\ell\geq 0} \mathbb{M}^{\ell}$ we denote the set of all finite words, which forms a semigroup under concatenation and with neutral element $\epsilon$. We focus on stationary processes, meaning that the probability of a sequence of letters
\begin{equation}
	p(\uu)\equiv \Pr\left\{\mathcal Y_t=u_1,\mathcal Y_{t+1}=u_2,\ldots,\mathcal Y_{t+\ell-1}=u_\ell\right\},\quad	\uu=(u_1,\ldots,u_\ell)\in\mathbb{M}^\ell
\end{equation}
does not depend on $t$. For the empty word, we have $p(\epsilon)=1$. The largest class of hidden cause models we consider is the class of quasi-realizations, defined as follows.

\begin{figure}[ht]
\centering
\includegraphics[width=15cm]{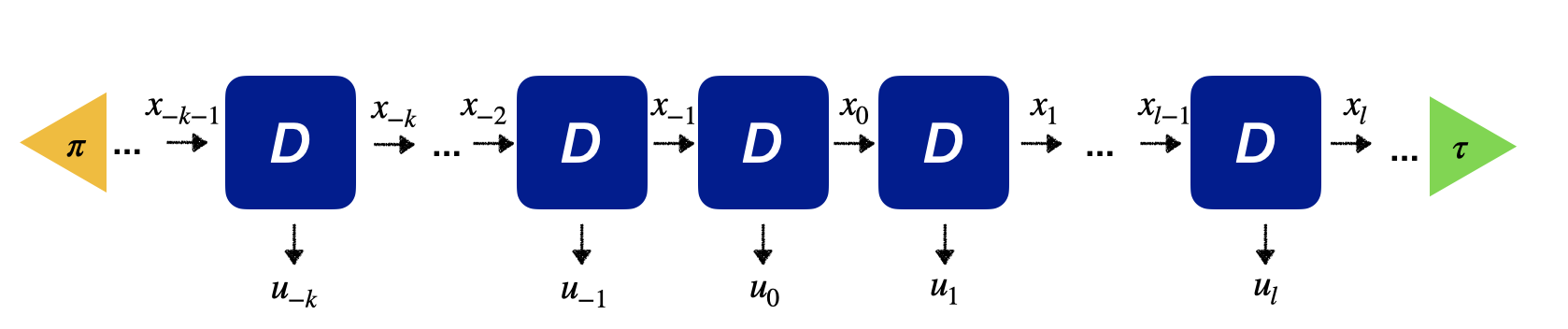}
\caption{A depiction of a general stationary process with finite memory. The probability of a sequence $u_{-k,..,u_0,...,u_l}$ can be computed as the inner product between a right stationary state $\pi$, evolved through a sequence of linear maps $D_{u_{-k}},..,D_{u_{l}}$ acting from the right, and a right stationary state $\tau$. The hidden vector space in which $\pi D_{u_{-k}},...D_{u_{l}}$ lives represents the memory of the process. For quantum hidden Markov models, $\pi$ is a state and $\tau$ is the trace functional in the dual of the state space, while $D_{u}$ are CP maps such that $\sum_{u\in \mathbb M} D_u$ is unital.}
\label{processfig}
\end{figure}

\begin{definition} 
A quasi-realization of a stationary stochastic process $p$ is a quadruple $(\V, \pi,$ $D,\tau)$,
where $\V$ is a real vector space, $\tau\in\V$, $\pi\in\V^*$, and 
$D:\mathbb{M}^*\rightarrow \mathcal L(\V)$ mapping a word $\uu\in\mathbb{M}^*$ to a linear map $D^{(\uu)}$ of $\V$ a semigroup homomorphism, i.e.
\begin{align}
\label{eq:Drepresentation}
	D^{(\emptyword)} = \operatorname{id}, \quad
	D^{(\uu)}D^{(\vv)} = D^{(\uu\vv)}\ \forall \uu,\vv\in \mathbb{M}^*.
\end{align}
In addition, the following fixed-point relations hold,
\begin{equation}
\label{eq:eigenrelations}
	\pi \left[\sum_{u\in\mathbb{M}}D^{(u)}\right] = \pi,
	\qquad
	\left[\sum_{u\in\mathbb{M}}D^{(u)}\right]\tau=\tau,
\end{equation}
\begin{equation}\label{quasi-realization}
    \text{and}\quad
	p(\uu)=\pi D^{(\uu)}\tau\quad \forall \uu\in\mathbb{M}^*.
\end{equation}
\end{definition}
 The right hand side of Eq.~(\ref{quasi-realization}) can be visually represented as in Fig.~\ref{processfig}.
Quasi-realizations that generate the same stochastic process are said to be {\bf equivalent}. Quasi-realizations of a process with minimal dimension of $\mathcal V$ are called regular, and they are related by each other by a similarity transformation, (i.e. for two equivalent regular realizations $(\V, \pi,$ $D,\tau)$, $(\V', \pi', D',\tau')$, $\V$ is linearly isomorphic to $\V'$ through an invertible linear map $T$, $\pi'=\pi T^{-1}$, $\tau'=T\tau$, $D'_{u}=TD'_{u}T^{-1}$. {Note that due to the semigroup law Eq.~(\ref{eq:Drepresentation}), $D$ is really given entirely by the maps $D^{(u)}$, $u\in\mathbb{M}$, making a quasi-realization a finite object in linear algebraic terms, as it can be given by a finite list of real numbers.}

The linear structure of quasi-realizations alone is not sufficient to guarantee the positivity of the probabilities. However, any quasi-realization of a stochastic process can be understood as arising from the dynamics of a (possibly exotic) general probabilistic theory. In fact, it is immediate to show that: 
\begin{proposition}
A quasi-realization defines a non-negative measure if and only if there 
is a convex cone $\C\subset\V$ such that $\tau\in\C$, $D^{(\uu)}(\C)\subseteq\C$, $\pi\in\C^* := \{ f\in\V^* : f(x) \geq 0 \ \forall x\in\C \}$, the dual cone of $\mathcal C$.
\end{proposition}
Note that, without loss of generality, the cone in the last proposition can be chosen to be closed: otherwise simply go to the closure of $\C$, $\overline{\C}=\C^{**}$, which is stable under the maps $D^{(\uu)}$ and has the same dual $\C^*$.
In fact, the cone $\mathcal C$ can be viewed as the cone of effects of a general probabilistic theory (GPT) with $\tau$ being the unit~\cite{Muller2021,Ludwig64,lami2018non}, and $\mathcal C^*$ as the cone of states. A pair of cones $\mathcal C$, $\mathcal C'\subseteq \mathcal C^{*}$ is what defines a general probabilistic theory; the maps $D^{(\uu)}$ stabilize the cone $\mathcal C$, and the ${D^{(\uu)}}^{\top}$ stabilize $\mathcal C^*$, therefore they can be considered as physical maps of the GPT.
A quasi-realization does not immediately identify a unique stable cone $\mathcal C$ in general. However, we can put inner and outer bounds on it from the cones generated by the quasi-realization dynamics itself.
\begin{proposition}\label{inclusions}
Any convex cone $\C\subset\V$ such that $\tau\in\C$, $D^{(\uu)}(\C)\subseteq\C$, $\pi\in\C^* := \{ f\in\V^* : f(x) \geq 0 \ \forall x\in\C \}$ has to satisfy the inclusions
\begin{equation}
\C_{\min}\subseteq \C\subseteq \C_{\max},
\end{equation}
where
\begin{align}
  \mathcal{C}_{\text{min}}
    &= \text{cone} \lbrace D^{(\mathbf{u})} \tau : \mathbf{u}\in \mathbb{M}^* \rbrace,\\ 
  \mathcal{C}_{\text{max}}
    & = \text{cone} \lbrace \pi D^{(\mathbf{u})}: \mathbf{u}\in \mathbb{M}^* \rbrace^*.
\end{align}
\end{proposition}

An important result in the theory of quasi-realizations is that a stochastic process has a finite-dimensional quasi-realizations if and only if the rank of a suitable Hankel-type matrix constructed from the probabilities of the finite words is finite.
This matrix $H$ is an infinite matrix with entries indexed by pairs of words, such that $H_{\uu,\vv}=p(\uu \vv)$. Writing the columns of $H$ as $h_{\vv}=H_{\cdot,\vv}$, a potentially infinite-dimensional quasi-realization in the column space $\mathcal{V} = \operatorname{span}\{h_{\vv}\}$ is obtained by choosing $\pi=(1,0,0,...) + \ker \mathcal{V}$, $\tau=h_{\epsilon}$ and $D^{(u)}h_{\vv}=h_{u\vv}$. This is a bona fide finite-dimensional quasi-realization if and only if the rank of $H$ is finite.
We will focus on such processes and denote their set as $\mathcal G$, with the idea in mind that they represent a privileged class of candidate processes, since they can in principle be reconstructed from a finite number of quantities, obtainable from observations of the process if enough data is available.

\subsection{Classical and quantum processes}
A subset $\mathcal P$ of $\mathcal G$ are those processes admitting a classical probability interpretation in finite dimension, denoted as \textbf{\emph{positive} realization}, also known as hidden Markov models. In this case the process $p$ admits a quasi-realization $(\mathbb{R}^d,\pi,D,\vec{1})$, such that $D^{(\uu)}$ are non-negative matrices and $\overline{D}=\sum_{u\in\mathbb{M}}D^{(u)}$ is (right) stochastic,
$\pi\in (\mathbb{R}^d)^*$ is a stationary distribution of $\overline{D}$, and $\vec{1}=(1,1,\ldots,1)\in \mathbb{R}^d$. A larger subsets is given by the processes $\mathcal C \mathcal P$ which admit a finite-dimensional quantum explanation, that is a \textbf{\emph{completely positive} realization}: in this case the quasi-realization can be chosen to be $(\BH^\sa,$ $\rho, D, \1)$, where $\BH$ is the space of bounded operators on some finite-dimensional Hilbert space $\H$ and $\BH^\sa$ the space of selfadjoint operators, $\rho$ is a positive semidefinite density operator in $\BH$, such that $D^{(\uu)}$ are completely positive maps on $\BH$ and $\overline{D}=\sum_{u\in\mathbb{M}}D^{(u)}$ is unital, and $\1$ is the identity of $\BH$. Positive and completely positive realization are guaranteed to give positive probabilities.

A natural question is then to ask if the inclusions $\mathcal P\subseteq \mathcal C\mathcal P\subseteq \mathcal G$ are strict. This question makes sense only if one restricts to finite memory systems, since from the infinite-dimensional quasi-realization we presented in the last paragraph, 
a HMM with countably infinite classical memory can be constructed~\cite{Vidyasagar2014,carlyle1967identification}. As already mentioned, $\mathcal P\subsetneq \mathcal G$ was shown as an early result by~\cite{fox1968, dharmadhikari1970}, while $\mathcal P\subsetneq \mathcal{CP}$ was shown first in~\cite{monras2016}. We are going to prove here that even $\mathcal{CP}\subsetneq \mathcal G$ holds. In order to show these separations, it is useful to establish necessary and sufficient conditions for a process to have a positive or completely positive realization. 

For the classical case, these were provided by~\cite{dharmadhikari1963sufficient}: Given a quasi-realization $(\V,\pi,D,\tau)$, an equivalent positive realization exists if and only if there is a convex pointed \textbf{polyhedral} cone $\C\subset \V$ such that
$\tau\in \C$, $D^{(v)}(\C)\subseteq \C$, $\pi\in\C^*$. For the quantum case, an analogous characterization was given in~\cite{monras2016} highlighting the role of \textbf{semidefinite representable cones}, defined as follows.

\begin{definition} 
Let $\V$ be a finite dimensional real vector space.
A \emph{\textbf{semidefinite representable (SDR) cone }} is a set $\C\subset\V$ such that there exists a subspace $\W\subseteq \B(\mathbb{C}^d)^\sa$ for some $d$ and a linear map $L:\W\rightarrow \V$ with
\begin{equation}\label{states}
	\C=L(\W^+),
\end{equation}
where $\W^{+}=\W\cap\PSD$, $\PSD$ being the cone of positive-semidefinite matrices. 
\end{definition}

For our purposes we will use that a necessary condition for a process to have a completely positive realization is that any regular representation of the same process must admit an SDR cone $\C\subset \V$ such that
$\tau\in \C$, $D^{(v)}(\C)\subseteq \C$, $\pi\in\C^*$ \cite{monras2016}.
Note that an SDR cone is \emph{semi-algebraic}, that is, it can be defined through a finite number of inequalities involving polynomials of the coordinates.

Since both the characterization of classical and of quantum processes do not give a prescription for how to find the stable polyhedral or SDR cone, respectively, they are not immediately usable to establish if a given process has a positive or completely positive realization. However, they are powerful enough to exclude the existence of such realizations if one is able to rule out the existence of stable cones with the desired properties.

\subsection{HMM vs HQMM}
The processes presented in~\cite{fox1968, dharmadhikari1970}, which we refer to as Fox-Rubin-Dharmadikari-Nadkarni (FRDN) processes, were shown to be in $\mathcal G$ by defining them explicitly as a function of Markov chains with infinite memory (non-negative integers as internal states), and then proving that the rank of the Hankel matrix $H$ is finite. As we have observed, this means that the processes can be explained with a finite-dimensional quasi-realization. 
In particular, the transition probabilities of the Markov chain are 
\begin{equation}\begin{split}\label{FRDN}
P(X_{i+1} =\ell | X_{i}=0) &= h_\ell, \quad \text{where } h_\ell := \lambda^\ell \sin^2 \left( {\ell\alpha}/{2} \right) \text{ for } \ell>0,\ h_0 := 1-{\textstyle{\sum_{\ell>0} h_\ell}} \\
P(X_{i+1} =\ell-1 | X_{i}=\ell) &=1 \text{ for } \ell>0,
\end{split}\end{equation}
and the function is defined as $f(0)=a$ and $f(x)=b$ if $x>0$, $\alpha\in \mathbb R$ and $0<\lambda\leq 1/2$. The resulting processes do not have a finite-dimensional classical realization when $\pi$ and $\alpha$ are not commensurate. It was unknown if the processes in~\cite{fox1968, dharmadhikari1970} had a quantum realization or not, and since the example was formulated naturally as an infinite-dimensional classical model, it could have been that it was sufficient to show the separation $\mathcal{CP} \subsetneq \mathcal G$. We show that this is not the case, since a quantum realization exists.
\begin{theorem}\label{qrel}
The process given by Eq.~(\ref{FRDN}) has a quantum realization on a qutrit, given by maps of the form
\begin{equation}
D^{\dagger}_{b}=\Pi_{01}\circ\Phi_{r,\alpha}\circ \Pi_{01},\quad 
D^{\dagger}_{a}(\rho)=\bigl(\Tr\rho-\Tr D_b^\dagger(\rho)\bigr) (p\ketbra{\xi}{\xi}+(1-p)\ketbra{2}{2}),
\end{equation}
where $\Pi_{01}$ is the cp map projecting onto $\operatorname{span}\{\ket{0},\ket{1}\}$, and $\Phi_{r,\alpha}$ is the qubit map
\begin{equation}
\Phi_{r,\alpha}(\rho)=\lambda e^{-r X}e^{i \alpha Z/2}e^{r X} \rho\, e^{r X}e^{-i \alpha Z/2}e^{-r X},
\end{equation}
for a suitable choice of $r\in \mathbb R$, $0\leq p\leq 1$ and $\ket{\xi}$ s.t. $\braket{\xi}{2}=0$, depending on $\alpha$ and $\lambda$.
\end{theorem}

To obtain this result, we first derive an explicit quasi-realization of the model (which was not given previously), and then looked for an equivalent quantum realization imitating the main features, in particular the eigenvalues of the maps. {Thus, the FRDN processes} cannot separate $\mathcal{CP}$ from $\mathcal G$.

Some remarks are in order:
\begin{itemize}
    \item The non-existence of a positive realization was proven by showing that in any realization the map $D_b$ must have eigenvalues with maximum modulus with arguments that are non-commensurate with $\pi$, which is impossible for nonnegative matrices by the Perron-Frobenius theorem~\cite{Johnson2017}.

    \item Theorem~\ref{qrel} defines bona fide HQMM even if $p$ and $\xi$ are not tuned to give exactly the FRDN models (only $r$ has to satisfy some constraints in order for $D_a^\dagger$ to be completely positive). The argument of the proof that there does not exist any finite-dimensional classical HMM implementing the process is unchanged, since the eigenvalues of the map $D_b$ do not change.

    \item The proof of the impossibility of a classical model for this family of quantum realizations differs somewhat from the argument provided for the family in~\cite{monras2016}, which defines processes that are naturally representable by a 2-qubit quantum systems, and the existence of a stable polyhedral cone was excluded directly by looking at the symmetry properties of the stable cones, which are incompatible with polyhedral cones. This approach of analyzing the problem geometrically proves to be decisive to prove the separation between quantum and general theories, as we will show in the next section. There, in fact, looking at spectra of the maps does not seem to help much.

\end{itemize}

When $\alpha$ is commensurate with $\pi$, say $\alpha/\pi=s/t$ with coprime integers $s$ and $t$, the {FRDN} models admit a positive (classical) realization, with a minimal dimension $t$~\cite{dharmadhikari1970}. In fact, when there are no eigenvalues with arguments incommensurate with $\pi$, the spectral argument cannot rule out classical realizations. However, the dimension of the minimal positive realization can be bounded from below, since the allowed region for eigenvalues of  $n\times n$ matrices with non-negative elements is a subset of the convex hull of the $k$-roots of unity, $k=1,\ldots,n$, multiplied by the maximum positive eigenvalue~\cite{Karpe1951}. We use this fact to prove a noise robustness results for the quantum processes of Theorem~\ref{qrel}, in presence of depolarizing noise, in the special case of $p=1$ where the process effectively take place on a qubit. 
We believe the argument can be adapted also for general $0\leq p<1$.

\begin{theorem}\label{theonoise}
For $0\leq q < 1$ and $0 < s\leq 1$, consider the processes defined by the HQMM with cp maps
\begin{align}
D_{b}^{\dagger}(q,s)  
  &=q\Phi_{r,\alpha}+(1-q)s\frac{\1}{2}\Tr \\ 
D^{\dagger}_{a}(\rho)(q,s)
  &=q\tr{(\1-\Phi_{r,\alpha}^{\dagger} (\1))\rho}\ketbra{\xi}{\xi}+(1-q)(1-s)\frac{\1}{2}\Tr,
\end{align}
at fixed $r\neq 0$ and varying $\alpha$.
If positive realizations exist for every $\alpha$, their maximum dimension (i.e. number of states of the HMM)  must be $\geq \Omega\left(\frac{\lambda}{s\sqrt{1-q}(\cosh 4r)}\right)$, assuming that $1-q$ is small enough.
\end{theorem}

\subsection{Processes without quantum realization}

Our main result is to present non semi-algebraic $3$-dimensional cones which are the only closed stable cones for models of certain stochastic processes, thus ruling out the possibility that these processes admit a quantum realization. These cones are defined as follows:

\begin{itemize}
\item Exponential cone:
\begin{equation}\label{eq:exponentialcone}
\mathcal{K}_{\exp} = \left\{ (x_1,x_2,x_3)\in\mathbb{R}^3 : \frac{x_1}{x_2} \geq e^{\frac{x_3}{x_2}}, x_2 > 0 \right\} \cup \{ (x_1,0,x_3) : x_1\geq 0 , x_3 \leq 0 \}.
\end{equation}

\item Power cones (for $0<\alpha<1$):
\begin{align}\label{eq:powercone}
\mathcal{K}_{\alpha} = \left\{ (x_1,x_2,x_3)\in\mathbb{R}^3 : x_1\geq 0,\, x_3\geq 0,\, x_1^{\alpha}x_3^{1-\alpha}\geq |x_2| \right\}.
\end{align}
\end{itemize}
Both $\mathcal{K}_{\exp}$ and the $\mathcal{K}_\alpha$ are closed convex cones, and they are all not semi-algebraic (the latter for irrational $\alpha$). Indeed, the boundary of $\mathcal{K}_{\exp}\cap\{x_2=1\}$ is the graph of the transcendental exponential function, $\{x_1=e^{x_3}\}$; likewise, the boundary of $\mathcal{K}_\alpha\cap\{x_2=1\}$ is the graph of the power function, $\{x_1 = x_3^{1-\frac{1}{\alpha}}\}$, which is transcendental for irrational $\alpha$.

The minimal example we can find, using an alphabet of $3$ letters, is the following.
\begin{theorem}\label{theoexp}
It is possible to choose $\nu,a,b\in \mathbb{R},m_0,\mu_0\in \mathbb{R}^3$, such that the linear maps: 
\begin{align}\label{eq:expquasirealization}
D_1 = \nu\begin{pmatrix}
 a & 0 & 0 \\
 0 & 1 & 0 \\
 0 & \ln a & 1
\end{pmatrix},
\quad
D_2 = \nu \begin{pmatrix}
 b & 0 & 0 \\
 0 & 1 & 0 \\
 0 & \ln b & 1
\end{pmatrix},
\quad
D_0 = \nu m_0 \mu_0^T, 
\end{align}
are such that $\overline{D}=D_0+D_1+D_2$ has unique left and right eigenvectors with eigenvalue $1$, respectively $\pi, \tau$, so that, with $D:\{0,1,2\}\rightarrow \mathcal L(\mathbb R^3)$ generated by ${D_0,D_1,D_2}$:
\begin{itemize}
    \item $(\mathbb R^3, \pi, D,\tau)$ is a bona fide regular quasi-realization of a stochastic process,
    \item $\mathcal{K}_{\exp}$ is the unique stable closed convex cone admitted by $(\mathbb R^3, \pi, D,\tau)$.
\end{itemize}
Thus, the resulting stochastic processes does not admit a quantum realization.
\end{theorem}


The crucial observation, as in \cite{monras2016}, is that any candidate closed stable cone $\mathcal C$ has to satisfy 
\begin{equation}
 \overline{\mathcal{C}_{\min}} = \overline{\operatorname{cone} \lbrace D^{(\mathbf{u})} \tau : \mathbf{u}\in \mathbb{M}^* \rbrace }\subseteq \mathcal C\subseteq \mathcal{C}_{\max} = \operatorname{cone} \lbrace \pi D^{(\mathbf{u})} :  \mathbf{u}\in \mathbb{M}^* \rbrace^*.
\end{equation}
On the other hand, for the given process {the parameters are chosen in such a way that} $\overline{\mathcal{C}_{\min}} = \mathcal{K}_{\exp} = \mathcal{C}_{\max}$, and therefore the only possible choice is $\mathcal C=\mathcal{K}_{\exp}$. Indeed the matrices are defined in such a way that after a reset, which must happen at some point, the rays generated by the repeated action of the matrices $D_1$ and $D_2$ in any order, densely explore the extremal rays of the exponential cone.

With the same strategy we can also show that also the power cones with irrational power give processes that are not representable by a HQMM. In this case the invertible matrices are diagonal, but we need an alphabet of four letters, rather than three.
\begin{theorem}\label{theopow}
It is possible to choose $\nu',a,b\in \mathbb{R},m_0,\mu_0\in \mathbb{R}^3$, such that the linear maps: 
\begin{align}\label{eq:powquasirealization}
D_1 &= \nu'\begin{pmatrix}
 a & 0 & 0 \\
 0 & 1 & 0 \\
 0 & 0 & a^{\frac{\alpha}{\alpha-1}}
\end{pmatrix},
\quad
D_2 = \nu' \begin{pmatrix}
 b & 0 & 0 \\
 0 & 1 & 0 \\
 0 & 0 & b^{\frac{\alpha}{\alpha-1}}
\end{pmatrix},
\\
D_3&= \nu'\begin{pmatrix}
 1 & 0 & 0 \\
 0 & -1 & 0 \\
 0 & 0 & 1
\end{pmatrix},
\quad
D_0 = \nu' m_0 \mu_0^T
\end{align}
are such that $\overline{D}=D_0+D_1+D_2+D_3$ has unique left and right eigenvectors with eigenvalue $1$, respectively $\pi, \tau$, so that, with $D:\{0,1,2,3\}\rightarrow \mathcal L(\mathbb R^3)$ generated by ${D_0,D_1,D_2,D_3}$:
\begin{itemize}
    \item $(\mathbb R^3, \pi, D,\tau)$ is a bona fide regular quasi-realization of a stochastic process,
    \item $\mathcal{K}_{\alpha}$ is the unique stable closed convex cone admitted by $(\mathbb R^3, \pi, D,\tau)$.
\end{itemize}
Thus, the resulting stochastic processes does not admit a quantum realization when $\alpha$ is irrational.
\end{theorem}

\begin{figure}
  \centering
  \includegraphics[width=.8\linewidth]{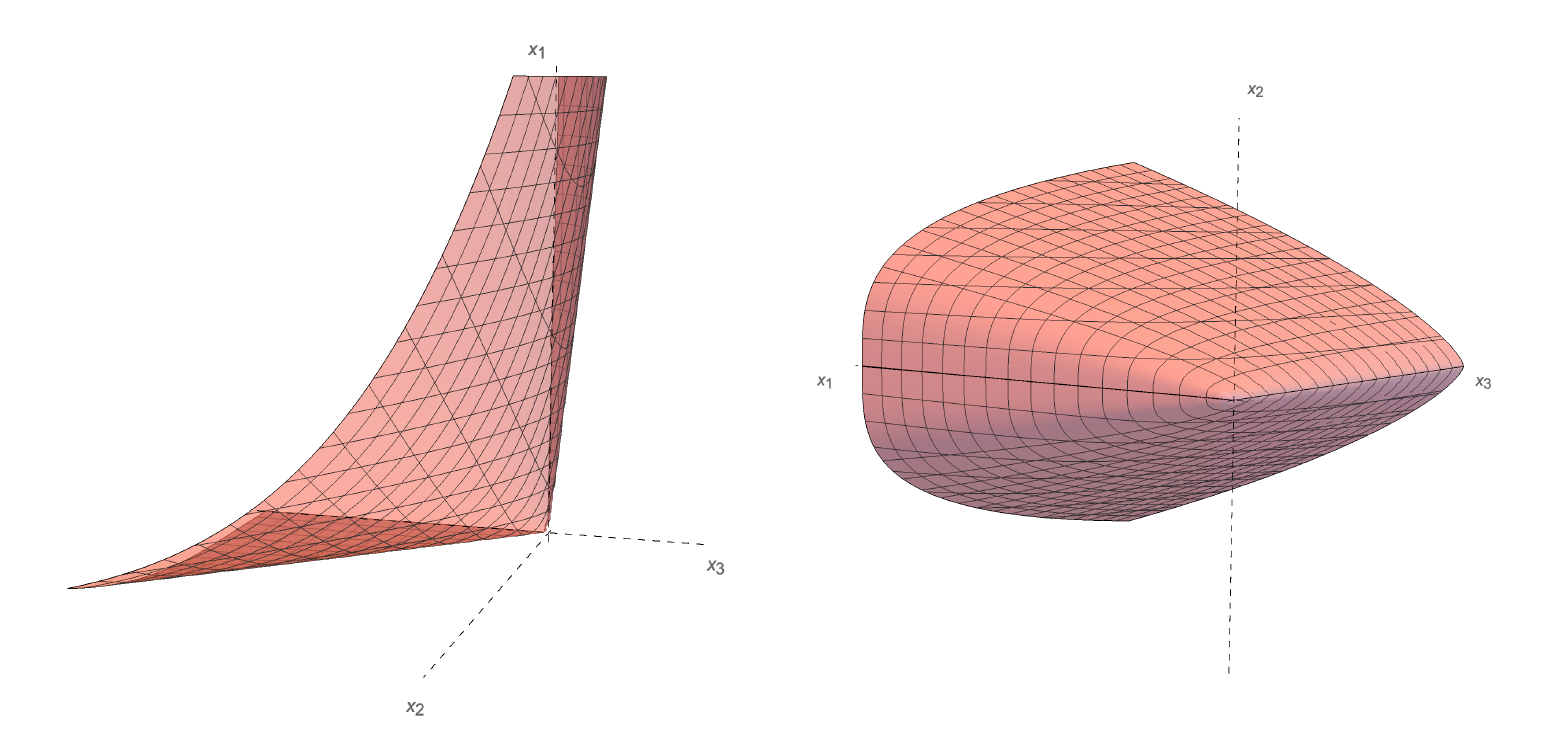}
\caption{Exponential cone~\eqref{eq:exponentialcone} (left) and power cone~\eqref{eq:powercone} (right) for $\alpha = \frac{1}{\sqrt{2}}$.}
\label{fig:cones}
\end{figure}

\section{Discussion}
\label{sec:discussion}
The result of the previous section has an important consequence: one could have hoped that $\mathcal C\mathcal P=\mathcal G$, meaning that quantum theory would be able to explain any sequence of observations from a finite GPT dynamics, and this property could be a principle that distinguishes quantum theory among general probabilistic theories. This is not the case, and the study of extensions of quantum mechanics giving rise to larger sets of quasi-realizations is interesting to pursue, with possible applications in data analysis applications, in many-body physics and in the foundations of quantum mechanics. In particular, the exponential and power cones discussed here, and their associated GPTs, have a rich symmetry structure, as indeed the respective cones are generated by the action of a group of matrices on their boundary, reminiscent of the fact that in quantum mechanics the pure states are the orbit of any fiducial pure state under the action of the unitary group. This translates into a large set of essentially reversible dynamics of the GPTs. 

As classical and quantum models are actually not restricted to a specific dimension, it is interesting to look for possible multivariate generalizations of power cones and exponential cones, which can be used to provide richer quasi-realizations, and which might unify classical, quantum and the present new state spaces (see e.g.~\cite{Fawzi2017}). Commutative multivariate generalizations that come to mind are ($\alpha\in\mathbb{R}^n$ with $\alpha_i \geq 0$ and $\sum_{i=1}^n \alpha_i = 1$): 
\begin{itemize}
 \item the multivariate power cone
 \begin{align}
        \mathcal{K}^n_{\alpha} = \left\{ (\vec{x},z)\in\mathbb{R}^{n}\times\mathbb{R} : x^\alpha_1  \cdots x^{\alpha_n}_n\geq |z|, x_i\geq 0 \right\};
 \end{align}

 \item and the multivariate exponential cone
 \begin{align}
    \mathcal{K}^n_{\alpha,\exp} = \left\{ (x,\vec{y},z)\in \mathbb{R}\times\mathbb{R}^n\times\mathbb{R} : \frac{1}{z} y_1^{\alpha_1}\cdots y_n^{\alpha_n} \geq e^{\frac{x}{z}} , y_i\geq 0 , z\geq 0 \right\}.
 \end{align}
\end{itemize}
These cones however can be represented with inequalities involving linear constraints and vectors belonging to the previously discussed $3$-dimensional exponential and power cones~\cite{chares2009cones}, therefore they are not really giving new structural building blocks.

On the other hand, and perhaps more interestingly from the point of view of quantum foundations, are extensions using positive semidefinite matrix cones, which reduce to the power cones and the exponential cones on specific sections, and to the positive semidefinite cone on others. As usual in non-commutative settings, there is more than one natural extension to matrices, and we briefly discuss a few possibilities.

\begin{itemize}
\item Matrix exponential cone: as the exponential function is not matrix convex nor monotone, we apply the logarithm (which is matrix monotone and concave), and define
\begin{equation}
    \mathcal{L}_{\exp} = \overline{\left\{ (A,B,t)\in \mathbb{R}^{d\times d} \times \mathbb{R}^{d\times d} \times\mathbb{R}: \log\frac{A}{t} \geq \frac{B}{t}, A > 0, B=B^\dagger, t > 0 \right\}}.
\end{equation}

\item There are at least two natural versions of matrix power cones (for $0<\alpha<1$ and a fixed $X\in\mathbb{R}^{d\times d}$), based on Lieb's concavity theorem:
\begin{align}
    \mathcal{L}_{\alpha,X} 
      &= \left\{ (A,B,t) \in \mathbb{R}^{d\times d} \times \mathbb{R}^{d\times d} \times \mathbb{R} : \Tr X^T A^\alpha X B^{1-\alpha} \geq |t|, A,B\geq 0 \right\}, \\
    \mathcal{L}_\alpha
      &= \left\{ (A,B,T) \in \mathbb{R}^{d\times d} \times \mathbb{R}^{d\times d} \times \mathbb{R}^{d^2\times d^2} : A^\alpha \otimes B^{1-\alpha} \geq T \geq 0, A,B\geq 0 \right\}, 
\end{align}
the latter admitting an obvious generalization to $\alpha_i\geq 0$, $\sum_{i=1}^n\alpha_i=1$ by way of $n$-fold tensor products. 

\item Matrix relative entropy cone
\begin{align}
    \mathcal{D} = \overline{\left\{ (A,B,t)\in \mathbb{S}^d_+\times \mathbb{S}^d_+ \times\mathbb{R} : \Tr(A\log A - A\log B) \leq t \right\}}.
\end{align}
\end{itemize}

Notice that the section $\{t=0\}$ of $\mathcal{L}_{\exp}$ is $\left\{(A,B) : A\geq 0, B\leq 0\right\}$. Both versions of the matrix power cone have the property that the section with $\{t=0\}$ (resp.~$\{T=0\}$) give just a double copy of the cone of positive semi-definite matrices in dimension $d$. Finally, $\mathcal{D}$ intersected with $\{t=0\}$ is $\left\{ (A,B)\in \mathbb{R}^{d\times d} \times \mathbb{R}^{d\times d} : A, B\geq 0, \supp A \subseteq \supp B,\Tr(A\log A - A\log B)=0\right\}$. This means that quantum dynamics can be obtained by projecting onto the $t=0, A=B$ hyperplane and applying the same CP map to $A$ and $B$. On the other hand, acting with the map which projects $A$ and $B$ to $(\Tr A)\1$ and $(\Tr B)\1$, and does not touch $t$, and then with the maps seen in the examples, one recovers the power cone and the exponential cone.
Transcendental matrix cones could be also useful in the study of finitely correlated states, and it would be interesting to exhibit genuinely quantum (e.g. not diagonal in a product basis as in our examples) finitely correlated states that are not C${}^*$-finitely correlated. 

Another important direction to investigate is the classical-quantum separation in the presence of noise, to understand to which extent classical models can simulate noisy dynamics. We have shown a specific example where the memory of the classical model has to increase as $\Omega\left((1-q)^{-\frac12}\right)$, where $q$ is the noise parameter and the noiseless case corresponds to $q=1$. This holds if we insists in looking for exact realizations, and it is likely to be a generic feature of quantum models without classical realizations. What happens if we allow some level of approximation has yet to be formalized and studied.

Finally, there is a lot of room for improvement of necessary and sufficient conditions for a process to have a quantum realization. It would be interesting to single out some criteria which are easily verifiable from a quasi-realization. For example, our proof for excluding a quantum realization is heavily based on the fact that there is only one possible stable cone, and its not SDR. In general the stable cone is not unique, and it would be interesting to find a way to exclude quantum realizations in this case. 

\section*{Acknowledgments}
The authors thank Alex Monr\`as for discussions on the power of HQMMs. 
%
%
JL is supported by the National Research Foundation, Prime Minister’s Office, Singapore and the Ministry of Education, Singapore under the Research Centres of Excellence programme.
MF and AW were supported by the Baidu Co.~Ltd.~collaborative project ``Learning of Quantum Hidden Markov Models'', and currently by the Spanish MINECO (project PID2019-107609GB-I00) with the support of FEDER funds, and the Generalitat de Catalunya (project 2017-SGR-1127). MF is also supported by a Juan de la Cierva Formaci\`on fellowship (Spanish MICIN project FJC2021-047404-I), with funding from MCIN/AEI/10.13039/501100011033 and European Union NextGenerationEU/PRTR. AW acknowledges furthermore support by the European Commission QuantERA grant ExTRaQT (Spanish MICIN project PCI2022-132965), by the Spanish MCIN with funding from European Union NextGenerationEU (PRTR-C17.I1) and the Generalitat de Catalunya, by the Alexander von Humboldt Foundation, and by the Institute for Advanced Study of the Technical University Munich.

\bibliography{biblio} 
\bibliographystyle{unsrt}


\begin{appendix}
\section*{Appendix}

The following sections contain the proofs of the results in the main body of the paper.

\section{Quantum realizations of the FRDN processes: proof of Theorem~\ref{qrel}}
Before proving the theorem, we first give a quasi-realization of the FRDN process~\cite{fox1968,dharmadhikari1970}. This allows us to write explicit expressions for the probabilities of words, and elucidates certain features inherited by the completely positive realization, which we then write down in the second step.

\subsection{Quasi-realization}



We present an explicit quasi-realization $(\mathcal{V},\pi , D ,\tau )$ of FRDN processes. We fix $\mathcal{V} = \mathbb{R}^4$ and 
\begin{align}
\tau = \begin{pmatrix}
1\\
1\\
1 \\
1
\end{pmatrix}.
\end{align}
The matrix corresponding to the output $b$ is defined as
\begin{align}
D_b := \lambda \begin{pmatrix}
 0 & 0 & 0 & 0 \\
 0 & 1 & 0 & 0 \\
 0 & 0 & \cos(\alpha) & \sin (\alpha ) \\
 0 & 0 & -\sin (\alpha ) & \cos (\alpha )
\end{pmatrix},
\end{align}
and the matrix corresponding to input $a$ is a rank one matrix defined as
\begin{align}
D_a := w\pi_0^{T},
\end{align}
for $\pi_0,w\in\mathbb R^{4}$ to be  determined.
We want to fix the vector $\pi_0\in\mathcal{V}$. In order to do that we consider the probabilities of the sequences $b^n = bb\ldots b$ after an $a$ is output, i.e.
\begin{align}\label{condprob}
p(b^n|a) = \sum_{l=n}^\infty h_l = \frac{\lambda^n}{4} \left( \frac{2}{1-\lambda} - \frac{(e^{i\alpha})^n}{1-\lambda e^{i\alpha}} - \frac{(e^{-i\alpha})^n}{1-\lambda e^{-i\alpha}}  \right). 
\end{align}
Using the above expression and $p(b^n|a) = \pi_0 D_b^n \tau$ for every $n\geq 0$ (this also fixes $(\pi_0\tau)=1$) we obtain the following:
\begin{align}
\pi_0^{T} = \begin{pmatrix}
1-\frac{\frac{2}{1-\lambda} - a_{\lambda,\alpha} - b_{\lambda,\alpha}}{4} &
\frac{1}{2(1-\lambda)}&
 -\frac{a_{\lambda,\alpha}}{4} &
\frac{-b_{\lambda,\alpha}}{4}
\end{pmatrix},
\end{align}
where $a_{\lambda,\alpha} $ and $b_{\lambda,\alpha}$ are defined as follows:
\begin{align}
a_{\lambda,\alpha} := \frac{1-\lambda\cos(\alpha)+\lambda\sin (\alpha )}{(1-\lambda\cos(\alpha ))^2+\lambda^2 \sin^2 (\alpha ) } \\
b_{\lambda,\alpha} :=  \frac{1-\lambda\cos(\alpha)-\lambda\sin (\alpha )}{(1-\lambda\cos(\alpha ))^2+\lambda^2 \sin^2 (\alpha ) }. 
\end{align}
Requiring that $(D_a+D_b)\tau=\tau$ we get 
\begin{align}
w = \begin{pmatrix}
1 \\  1 - \lambda \\ 1 + \lambda \left( \sin (\alpha ) - \cos (\alpha ) \right) \\ 1-\lambda\left( \sin(\alpha ) + \cos (\alpha ) \right)
\end{pmatrix}.
\end{align}
This construction fully determines also $p(b^n a |a)$, therefore all the probabilities $p(\mathbf{u}|a)$. 

We are left with checking that $p(a)$ is positive and equal to the desired value.

We have the condition $\pi D_a+\pi D_b =\pi$, which implies
\begin{equation}
(\pi w)\pi_0^{T} (\1-D_b)^{-1}=\pi,
\end{equation}
since $\1-D_b$ is invertible.

Now, the candidate left fixed point $\pi$ satisfies
\begin{equation}
p(a)= \pi D_a \tau = (\pi w)= [\pi_0^{T}(\1-D_b)^{-1}\tau]^{-1}=\left[1+\sum_{l=1}^{\infty}h_l\right]^{-1},
\end{equation} 
which is the desired value. By virtue of the fixed point cosntraints and of the reset property, the probabilities of all words are completely determined and they coincide with those given by the FRDN process.

\subsection{Quantum (completely positive) realization}\label{sec:theorem2}

We are going to verify that the quantum process given in Theorem~\ref{qrel} gives the probabilities of the FRDN process. 

To start, observe that $\Phi_{r,\alpha}$ is a completely positive map and that its non-zero eigenvalues coincide with those of $D_b$. Recall 
\begin{equation}
\Phi_{r,\alpha}^n(\rho)=\lambda^n e^{-r X}e^{i n \alpha Z/2}e^{r X} \rho\, e^{r X}e^{-i n \alpha Z/2}e^{-r X}.
\end{equation}
Now, defining $\ket{\psi}=e^{-r X}\ket{0}$, $\ket{\phi}=e^{-r X}\ket{1}$, the eigenvalues and eigenvectors of $\Phi_{r,\alpha}^n$ are
\begin{align}
\Phi_{r,\alpha}^n(\ketbra{\psi}{\psi}) &=\lambda^n\ketbra{\psi}{\psi},\\
\Phi_{r,\alpha}^n(\ketbra{\psi}{\phi}) &=\lambda^ne^{i n\alpha}\ketbra{\psi}{\phi},\\
\Phi_{r,\alpha}^n(\ketbra{\phi}{\psi}) &=\lambda^ne^{-i n\alpha}\ketbra{\phi}{\psi},\\
\Phi_{r,\alpha}^n(\ketbra{\phi}{\phi}) &=\lambda^n\ketbra{\phi}{\phi}.\\
\end{align}
We set $\ket{\xi}=\frac{e^{i\phi}}{\sqrt{2}}\ket{0}+\frac{e^{-i\phi}}{\sqrt{2}}\ket{1}=\beta\ket{\psi}+\gamma\ket{\phi}$, with
\begin{align}
\beta\braket{0}{\psi}+\gamma\braket{0}{\phi} &=\beta \cosh{r}-\gamma \sinh{r} =\frac{e^{i\phi}}{\sqrt{2}},\\
\beta\braket{1}{\psi}+\gamma\braket{1}{\phi} &=-\beta \sinh{r}+\gamma\cosh{r}=\frac{e^{-i\phi}}{\sqrt{2}},
\end{align}
therefore
\begin{align}
\beta &=\frac{1}{\sqrt{2}} \frac{e^{i\phi}\cosh{r}+e^{-i\phi}\sinh{r}}{\cosh^2{r}-\sinh^2{r}}=\frac{e^{i\phi}\cosh{r}+e^{-i\phi}\sinh{r}}{\sqrt{2}}, \\
\gamma &=\frac{1}{\sqrt{2}} \frac{e^{-i\phi}\cosh{r}+e^{i\phi}\sinh{r}}{\cosh^2{r}-\sinh^2{r}}=\frac{e^{-i\phi}\cosh{r}+e^{i\phi}\sinh{r}}{\sqrt{2}}. 
\end{align}

We must have
\begin{align}p(b^n|a)=
\Tr[\Phi_{r,\alpha}^n p \ketbra{\xi}{\xi}]&=\Tr[\Phi_{r,\alpha}^n \left( |\beta|^2\ketbra{\psi}{\psi}+|\gamma|^2\ketbra{\phi}{\phi}+\overline{\gamma}\beta\ketbra{\psi}{\phi}+\overline{\beta}\gamma\ketbra{\phi}{\psi} \right)]\nonumber\\
&= \lambda^n (p( |\beta|^2+ |\gamma|^2)\cosh{2r}-p\overline{\gamma}\beta\sinh{2r}e^{in\alpha}-p\overline{\beta}\gamma\sinh{2r}e^{-in\alpha}).
\end{align}

In order to be compatible with Eq.~(\ref{condprob}), we thus need
\begin{align}
p(|\beta|^2+|\gamma|^2)\cosh{2r} &=\frac{1}{2(1-\lambda)}, \label{cond1}\\
p\overline{\gamma}\beta\sinh{2r} &= \frac{1}{4(1-\lambda e^{i\alpha})}, \label{cond2}
\end{align}
and we note that $\beta=\overline{\gamma}$, therefore we obtain, imposing $r\geq 0$

\begin{equation}\label{tanh}
\tanh{2r}=\frac{(1-\lambda)}{|1-\lambda e^{i\alpha}|},
\end{equation}
which is less than $1$ if $0<\lambda\leq 1/2$, and
\begin{equation}
\arg \beta=\arctan e^{-2r}\tan \phi=\frac{1}{2}\arctan\frac{\lambda\sin{\alpha}}{1-\lambda\cos{\alpha}},
\end{equation}
which has as a solution $\tan \phi= e^{2r}\tan\left(\frac{1}{2}\arctan \frac{\lambda\sin{\alpha}}{1-\lambda\cos{\alpha}}\right)$, and the expression for $\arg \beta$ comes from
\begin{align}
\sqrt{2}\beta=\cos{\phi}(\cosh{r}+\sinh{r})+i\sin{\phi}(\cosh{r-\sinh{r}})&=e^r\cos{\phi}+i e^{-r} \sin{\phi}\\&=\sqrt{e^{2r}\cos^2{\phi}+e^{-2r}\sin^2{\phi}}e^{i \arctan e^{-2r}\tan \phi}.
\end{align}

With this choice of $r$ and $\phi$, we have that the value of $p$ that solves Eq.~(\ref{cond1}) and~(\ref{cond2}) is the same. To compute it, observe that
\begin{equation}
(|\beta|^2+|\gamma|^2)\cosh{2r}-(\overline{\gamma}\beta+\overline{\beta}\gamma)\sinh{2r}=1,
\end{equation}
therefore we get
\begin{equation}
p=\frac{1}{2(1-\lambda)}-\frac{1}{4(1-\lambda e^{-i\alpha})}-\frac{1}{4(1-\lambda e^{i\alpha})}=\frac{1}{2(1-\lambda)}-\frac{1-\lambda\cos{\alpha}}{2(1+\lambda^2-2\lambda\cos\alpha)}.
\end{equation}
Note that
\begin{align}
\frac{1}{2(1-\lambda)}\geq\frac{1}{2(1-\lambda)}-\frac{1-\lambda\cos{\alpha}}{2(1+\lambda^2-2\lambda\cos\alpha)}\geq \frac{1}{2(1-\lambda)}-\frac{1}{2(1-\lambda)}=0
\end{align}
since
\begin{equation}
(1-\lambda\cos\alpha)(1-\lambda)=1+\lambda^2-\lambda(1+\cos\alpha)\leq (1+\lambda^2-2\lambda\cos\alpha),
\end{equation}
therefore $0\leq p\leq 1$ as desired.

We also need to check that $\Phi_{r,\alpha}(\rho)$ is trace non-increasing, that is
\begin{equation}
\Phi_{r,\alpha}^{\dagger}(\1)=\lambda e^{r X}e^{-i \alpha Z/2}e^{-2r X}e^{i \alpha Z/2}e^{r X}\leq \1,
\end{equation}
which is guaranteed since the eigenvalues of $\Phi_{r,\alpha}^{\dagger}(\1)$ are
\begin{align}
\omega_{\pm}&=\lambda\cosh^2{2r}-\cos{\alpha}\sinh^2{2r}\pm\frac{1}{2}\sqrt{\sinh^2{2r}(5+3\cosh{4r}+2\cos{\alpha}\sinh^2{2r})-2\cos{\alpha}\sinh^2{4r}},
\end{align}
which evaluate to $\omega_{+}=1$ and $\omega_{-}=\lambda^2$ when we substitute the value of $r$ given by Eq.~(\ref{tanh}). Finally, $p(a)$ is fixed as in the quasi-realization.

\section{Noise robustness of the size of classical memory: proof of Theorem~\ref{theonoise}}
The impossibility of classical realization fo FRDN models crucially use the fact that the maps have eigenvalues with phases which are not powers of roots of unity. This cannot happen for irreducible maps~\cite{wolf2012quantum}. Taking the qubit reduction of our example quantum realization (just take $(p=1)$ and choose the initial state to be in the $\{\ket{0},\ket{1}\}$ subspace), mixing our invertible map with completely depolarizing noise, say
\begin{align}
D_{b}^{\dagger}(q,s)&=q\Phi_{r,\alpha}+(1-q)s \frac{\1}{2}\Tr, \\ D^{\dagger}_{a}(q,s)&=q\tr{(\1-\Phi_{r,\alpha}^{\dagger} (\1))(\cdot)}\ketbra{\xi}{\xi}+(1-q)(1-s) \frac{\1}{2}\Tr,
\end{align}
when $q\neq1, s\neq0$, and the maximum modulus eigenvalues of $D_{b}^{\dagger}(q,s)$ have phases that are commensurate with $\pi$, since $D_{b}^{\dagger}(q,s)$ is irreducible~\cite{wolf2012quantum}. Classical realizations cannot be excluded in this way, but it is interesting to understand how large the dimension of the memory should be as $q$ approaches one, and this can be understood again looking at eigenvalues.  

In fact we have that
\begin{align}\label{pole}
\sum_{n=0}^{\infty}z^{-n} p(ab^na)&=\sum_{n=0}^{\infty}z^{-n} \pi D_a(q,s) D_{b}^{n}(q,s) D_a(q,s) \tau = \pi D_a(q,s) (\1-z^{-1} D_b)^{-1} D_a(q,s) \tau\\&=z^{-1}\pi D_a(q,s) (z \1 -D_b)^{-1} D_a(q,s) \tau.
\end{align}

These relation hold for any quasi-realization{, for every value of $1/z$ inside the radius of convergence of $\sum_{n=0}^{\infty}z^{-n}D_{b}^{n}(q,s)=:f(1/z)$, i.e. $|1/z|\leq ||D_{b}(q,s)||$, with $||D_{b}(q,s)||$ being the operator norm. This holds} in particular if the quasi-realization is classical. { From the quantum realization one obtains that a meromorphic continuation of $f(1/z)$ on all $\mathbb{C}$, since $f(1/z)$ is rational; by inspection, the continuation can have poles only for $1/z=1/\lambda$, where $\lambda$ is an eigenvalue of $D_{b}(q,s)$. Any classical realization will result in a function of $1/z$ coinciding with the function obtained from the quantum realization inside the minimum radius of convergence, therefore resulting in the same meromorphic continuation. We note that, again by inspection, the meromorphic continuation for a given quasi-realization has poles only at $z=\lambda$, where $\lambda$ is an eigenvalue of $D_b(q,s)$, and thus if a pole at $\lambda$ exists for the meromorphic continuation of the quantum realization, $\lambda$ has to be an eigenvalue of $D_{b}(q,s)$ in any realization. }

For $n\times n$ non-negative stochastic matrices, the allowed region of the eigenvalues is contained in the convex hull of $k$-roots of unity, $k\leq n$~\cite{Johnson2017,Karpe1951}, and this holds also for general non-negative matrices once their maximum eigenvalue is renormalized to one, since they are similar to a stochastic one~\cite{Johnson1981}. We can thus determine a lower bound on the dimension of the classical memory by showing that there are eigenvalues of the quantum map $D_b(q,s)$, associated to poles in Eq.~(\ref{pole}), which are outside the allowed region unless $n$ is large enough.
Suppose that two eigenvalues of $D_b^{\dagger}(1,0)$ are $\eta_{max}$ (which is on the maximal circle and real) and $\eta$. 
First of all, we observe that a perturbation bound constrains the eigenvalues of $D_b^{\dagger}(q,s)$ for $q\neq 1$. First of all, $D_b^{\dagger}(q,s)$ is similar (similarity of matrices $R,R'$ here means $R'=SRS^{-1}$ for some invertible matrices $S$) to $e^{rX}D_b^{\dagger}[e^{-rX} \cdot e^{-rX}]e^{rX}=q\lambda e^{i\alpha Z/2}\rho e^{-i\alpha Z/2}+(1-q)s\Tr[e^{-2rX}\cdot ]\frac{e^{2rX}}{2}$. Let $\eta'_{\max}$ be its maximum modulus eigenvalue, which is real and positive. $q\lambda  e^{i\alpha Z/2}\rho  e^{-i\alpha Z/2}+(1-q)s\Tr[e^{-2rX}\cdot ]\frac{e^{2rX}}{2}$ has an eigenvector $\ketbra{0}{0}-\ketbra{1}{1}$ with eigenvalue $q\lambda$, therefore $\eta'_{\max}\geq q\lambda$. We denote $\sigma(A)$ the $n$-tuple of eigenvalues of the $n\times n$ matrix $A$, counted with algebraic multiplicity. The optimal matching distance between two $n$-tuples $u,v$ is $d(u,v)=\min_{g \, \text{permutation}}\max_{1\leq i\leq n}|u_i-v_{g(i)}|$. Theorem VI.5.1 in~\cite{bhatia2013matrix} says that for a normal matrix $A$ and an arbitrary matrix $B$ such that $||A-B||$ is less than half the distance between any two distinct eigenvalues of $A$, then $d(\sigma(A),\sigma(B))\leq||A-B||$. In our case, the eigenvalues of $A=q\lambda U\cdot U^{\dagger}$, where $U=e^{i\alpha Z/2}$, are $\{q\lambda,q\lambda, q\lambda e^{i\alpha},q\lambda e^{-i\alpha}\}$, and the half the minimum distance between distinct eigenvalues is more than $q\lambda |\sin(\alpha)|$. By taking $B=A+(1-q)s\Tr[e^{-2rX}\cdot ]\frac{e^{2rX}}{2}$ we have that $||A-B||=(1-q)s 2\cosh(4r).$ Denoting $\{\eta_i\}$ and $\{\eta'_i\}$ the eigenvalues of respectively $A$ and $B$, note also that $d(\sigma(A),\sigma(B))\geq \min_{g \, \text{permutation}}|\eta'_i-\eta_{g(i)}|$ for any $i$, and that $\min_{i=1,..,4}|\eta'_{max}-\eta_{i}|=|\eta'_{max}-\eta_{max}|$. Supposing that $q$ is such that $||A-B||\leq q\lambda |\sin(\alpha)|$, we can find $|\eta'_{max}-\eta_{max}|\leq d(\sigma(A),\sigma(B))\leq (1-q)s 2\cosh(4r)$ and also an eigenvalue $\eta'$ such that $|\eta'-\eta|\leq d(\sigma(A),\sigma(B))\leq (1-q)s 2\cosh(4r)$. 

By repeated application of the triangle inequality, and supposing $2(1-q)s\cosh(4r)\leq q\lambda|\sin\alpha|$, we have the following:
\begin{align}
\label{eigcircle}
 \left|1-\frac{\eta}{\eta_{\text{max}}}-(1-\frac{\eta'}{\eta'_{\text{max}}})\right| &=\left|-\frac{\eta}{\eta_{\text{max}}}+\frac{\eta'}{\eta'_{\text{max}}}\right|=\left|-\frac{\eta-\eta'}{\eta'_{\text{max}}}+\eta\left(\frac{1}{\eta_{\text{max}}}-\frac{1}{\eta'_{\text{max}}}\right)\right|\nonumber\\
&\leq \frac{4(1-q)s\cosh(4r)}{q\lambda}.
\end{align}

Let us focus on the segment between $1$ and $e^{i\frac{2\pi}{n}}$: if $(1-\frac{\eta'}{\eta'_{\text{max}}})$ is outside the bigger circular segment individuated by the segment, then there is no classical model with such eigenvalues in dimension $n$, because this point is outside the convex hull of ${e^{ir\pi/k}, r=0,...,k-1, k=0,...,n}$. The maximum distance between this segment and the boundary of the circle is $1-\cos(\pi/n)$, which happens at $\alpha=\pi/n$. For this value of $\alpha$ there is not a classical model of memory smaller than $n$ if $2(1-q)s\cosh(4r)\leq q\lambda|\sin(\pi/n)|$ and $4(1-q)s\cosh(4r)\leq q\lambda(1-\cos(\pi/n))$ from Eq.~(\ref{eigcircle}). Since $|\sin(\pi/n)|\geq (1-\cos(\pi/n))/2\geq\frac{1}{6}(\pi/n)^2$, it is sufficient to require $4(1-q)s\cosh(4r)\leq q\lambda\frac{1}{6}(\pi/n)^2$ to exclude the existence of a classical model. Therefore if a classical model exists we need $\frac{1}{6}(\pi/n)^2<  \frac{ 4(1-q)s\cosh(4r)}{q\lambda}$.

We now have to show that in fact there are poles of $f(1/z)$ corresponding to $\eta'_{\max}$ and $\eta'$. Since probabilities are real, if a complex eigenvalue is a pole, its conjugate must be too. We also note that in our example $D_b(q,s)$ is guaranteed diagonalizable if $2(1-q)s\cosh(4r)\leq q\lambda|\sin\alpha|$. In fact, that this map is completely positive, therefore it admits a positive semi-definite eigenvector with real eigenvalue. We note that the operator $e^{rX}(\ketbra{0}{0}-\ketbra{1}{1})e^{rX}$ is an eigenvector with eigenvalue $q\lambda$, therefore a linear independent eigenvector with real eigenvalue exists. Finally, for these values of $q$, $D_b^{\dagger}(q,s)$ admits two distinct complex eigenvalues, again by $d(\sigma(A),\sigma(B))\leq||A-B||$. Since $D_b(q,s)$ is a $4\times 4$ matrix, it must be diagonalizable. 
This implies that if a complex eigenvalue $\eta'$ is not a pole, it means that either $D_a(q,s)\tau=0$ or $\pi D_a(q,s)=0$, which is excluded 
by looking at the definition of $D_a(q,s)$ for $q\neq 1, s\neq 1$, or that $D_a(q,s)\tau$ is orthogonal to some the right eigenspace of $D_b(q,s)$ corresponding to $\eta'$, or $\pi D_a(q,s)$ is orthogonal to some left eigenspace of $D_b(q,s)$ corresponding to $\eta'$. The latter two conditions are excluded by observing that the span of the orbits 
$\mathrm{span}\{D_b(q,s)^n D_a(q,s)\tau,n\geq 0\}$, $\mathrm{span}\{\pi D_a(q,s)D_b(q,s)^n,n\geq 0\}$, are at least 3-dimensional (therefore both complex eigenvalue are poles). 
This is seen explicitly for $q=1$, and for other values one can observe that the orbit is generated by linear combinations of the vectors in the orbit of the case $q=1$ and $\1$, in both cases. Since the orbits for $q=1$ densely explore a cone which is a linear transformation of a circular cone, there are always at least two points on the cone such that $\1$ is not in their span, therefore also in the case $q\neq1$ the orbits must span at least a three dimensional space.

\section{Processes without a quantum realization}
In this section we prove that there exist stochastic processes with a finite dimensional quasi-realization and that are not quantum realizable.

\subsection{Proof of Theorem~\ref{theoexp}: Exponential cone }\label{sec:theorem3}

Recall the definition of the exponential cone:
\begin{align}\label{eq:exponentialcone:app}
\mathcal{K}_{\exp} = \lbrace (x_1,x_2,x_3)\in\mathbb{R}^3 : x_1 \geq x_2 e^{\frac{x_3}{x_2}}, x_2 >0 \rbrace \cup \lbrace (x_1,0,x_3) : x_1\geq 0 , x_3 \leq 0 \rbrace.
\end{align}
We consider a quasi-realization on $\mathcal{V} = \mathbb{R}^3$, alphabet $\mathbb{M} = \lbrace 0,1,2 \rbrace$ and generators
\begin{align}\label{eq:expquasirealization:app}
D_1 = \nu\begin{pmatrix}
 a & 0 & 0 \\
 0 & 1 & 0 \\
 0 & \ln a & 1
\end{pmatrix},
\quad
D_2 = \nu \begin{pmatrix}
 b & 0 & 0 \\
 0 & 1 & 0 \\
 0 & \ln b & 1
\end{pmatrix},
\quad
D_0 = \nu m_0 \mu_0^T,
\end{align}
where $a>1>b>0$, $a+b\neq 2$, $\frac{\ln(a)}{\ln (b)} \in \mathbb{R}\setminus\mathbb{Q}$ (incommensurate), and 
\begin{align}
m_0 =
\begin{pmatrix}
m_{01} \\
m_{02} \\
m_{03}
\end{pmatrix} \quad
\mu^T_0 =
\begin{pmatrix}
\mu_{01} & \mu_{02} & \mu_{03}
\end{pmatrix}.
\end{align}

Here, $\nu$ is a normalization constant such that the largest absolute value of the (in general complex) eigenvalues of $D_0+D_1+D_2$ is $1$.

In order to check that the above quasi-realization defines a non-negative measure we are going to use a standard result that states this happens if and only if there is a convex cone $\mathcal{C}\subset\V$ such that $\tau\in\mathcal{C}$, $D^{(\uu)}(\mathcal{C})\subseteq\mathcal{C}$, $\pi\in\mathcal{C}^* = \{ f\in\V^* : f(x) \geq 0 \ \forall x\in\mathcal{C} \}$. Thus we need to describe what kind of cone $\mathcal{C}$ is preserved under the transformations $\lbrace D^{(\mathbf{u})} \rbrace_{\mathbf{u}\in\mathbb{M}}$. In fact, we argue that for any non zero stable convex cone $\mathcal{C}$ under all the transformations $D_u$ we can find $\tau\in \mathcal{C}$ such that $\left[ \sum_{u\in\mathbb{M}} D_u \right] \tau = \tau$.  This is a consequence of a generalized version of Perron-Frobenius theorem~\cite{brundu2018cones, vandergraft1968spectral,Yoshida_2020} that states that if $K$ is a convex cone preserved by a nonzero matrix $A$ then:
\begin{itemize}
    \item The spectral radius $\rho(A)$ is an eigenvalue of $A$.
    \item The cone $K$ contains an eigenvector of $A$ corresponding to $\rho(A)$.
\end{itemize}

It can be shown by inspection that $D_1,D_2$ preserve $\mathcal{K}_{\exp}$ acting from the left on column vectors, and $D_{0}$ also does it provided that we choose $\mu_0\in \mathcal{K}_{\exp}^*$ and $m_0\in \mathcal{K}_{\exp}$, therefore one can find $\nu>0$ such that $(D_0+D_1+D_2)\tau=\tau$, $\tau\in \mathcal{K}_{\exp}$. The same argument can be applied to $D_0,D_1,D_2$ acting from the right on row vectors, which preserve $\mathcal{K}_{\exp}^*$, therefore there exists $\pi\in \mathcal{K}_{\exp}^*$ such that $\pi(D_0+D_1+D_2)=\pi$.



The minimal stable cone is given by

\begin{align}\label{eq:coneminexp}
\mathcal{C}_{\text{min}} = \text{cone} \lbrace D^{(\mathbf{u})} \tau : \mathbf{u}\in \mathbb{M}^* \rbrace,
\end{align}
and what we just observed shows that $\C_{\min}\subseteq \mathcal{K}_{\exp}$.

On the other hand, provided that $D_0\tau\neq0$, we also have
\begin{align}\label{eq:coneminexp2}
\overline{\mathcal{C}_{\min}} \supseteq \overline{\text{cone} \lbrace \nu^{-s-t}D_1^s D_2^t m_0 : s\in \mathbb{N},t\in\mathbb{N} \rbrace }.
\end{align}

Indeed, when exploring the dynamics of this quasi-realization the operator $D_0$ acts as a "reset" to $m_0$ since it is defined as a rank-1 projector. We can ensure that $D_0\tau\neq 0$ in the following way.

If $D_0 \tau = 0$ then if $\ln (ab) \neq 0$ (which is true by the incommensurate condition) and $a+b\neq 2$ we have by definition of $\tau$ that, defining $e_1:=\begin{pmatrix}1\\0\\0\end{pmatrix}$ and $e_3:=\begin{pmatrix}0\\0\\1\end{pmatrix}$, we get 
\begin{align}
    \tau \in \text{span} \{e_1\} \qquad\text{or}\qquad  \tau \in \text{span} \{e_3\}.
\end{align}

If we now fix $D_0$ such that $D_0e_1\neq 0,D_0e_3\neq 0$ (simply choose $\mu_{01}\neq 0$, $\mu_{03}\neq 0$), then we have a contradiction.

Looking back at the orbit of $m_0$, The matrices $D_1$ and $D_2$ commute, so it suffices to consider
\begin{align}\label{eq:generator}
\nu^{-s-t}D_1^s D_2^t = \begin{pmatrix}
e^x & 0 & 0\\
0 & 1 & 0 \\
0 & x & 1
\end{pmatrix},
\end{align}
where $x = s\ln ( a)+t\ln (b )$. Note that $x\in\mathbb{R}$ is dense due to the incommensurability condition and Kronecker's Theorem. Thus, 
\begin{equation}
\overline{\text{cone}\left\{ \nu^{-s-t}D_1^s D_2^t m_0 : s\in \mathbb{N},t\in\mathbb{N}\right\}}  = \overline{\text{cone}\left\{  \begin{pmatrix}
e^x & 0 & 0\\
0 & 1 & 0 \\
0 & x & 1
\end{pmatrix}m_0 : x\in \mathbb{R}\right\}}.
\end{equation}
It is easy to see that 
\begin{align}\label{eq:epigraph}
 t\geq e^x \iff (t,1,x)\in \mathcal{K}_{\exp},   
\end{align}
or in other words that the epigraph $t\geq e^x$ is a section of $\mathcal{K}_{\exp}$. Setting $m_{02}=1$ and $m_{01}=e^{m_{03}}$ we thus have that the orbit of $\tau$ densely explores the curve $(e^x,1,x)$, and its closed conic hull is 
\begin{equation}
\overline{\text{cone} \lbrace \nu^{-s-t}D_1^s D_2^t m_0 : s\in \mathbb{N},t\in\mathbb{N} \rbrace }=\overline{\text{cone} \lbrace(e^x,1,x),x\in\mathbb R\rbrace}=\mathcal{K}_{\exp}.
\end{equation}
This can be seen as follows: 
\begin{itemize}
\item
$\mathcal{K}_{\exp}=\overline{\lbrace (x_1,x_2,x_3)\in\mathbb{R}^3 : x_1 \geq x_2 e^{\frac{x_3}{x_2}}, x_2 >0 \rbrace}$~\cite{chares2009cones}, and 
\item $\text{int}\left(\mathcal{K}_{\exp}\right):=\lbrace (x_1,x_2,x_3)\in\mathbb{R}^3 : x_1 \geq x_2 e^{\frac{x_3}{x_2}}, x_2 >0 \rbrace=\text{cone} \lbrace(e^x,1,x),x\in\mathbb R\rbrace$,
since for any $(x_1,x_2,x_3)\in \text{int}\left(\mathcal{K}_{\exp}\right)$,  $(\frac{x_1}{x_2},1,\frac{x_3}{x_2})$ is contained in the convex hull of $\lbrace(e^x,1,x),x\in\mathbb{R}\rbrace$ by convexity of the exponential function, and thus convexity of its epigraph (as a set).
\end{itemize}
This means that
\begin{align}\label{eq:coneminmax}
\overline{\mathcal{C}_{\min}}=\mathcal{K}_{\exp}.
\end{align}

The dual of $\mathcal{K}_{\exp}$ is given by
\begin{align}\label{eq:dualexponentialcone}
\text{int}\left(\mathcal{K}_{\exp}^*\right) &:=\text{cone} \left\lbrace (y_1,y_2,y_3)\in\mathbb{R}^3 :  y_1 \geq -y_3 e^{\frac{y_2}{y_3}-1}, y_1 > 0 , y_3 < 0 \right\rbrace\\
    \mathcal{K}_{\exp}^* &= \overline{\text{int}\left(\mathcal{K}_{\exp}^*\right)}.
\end{align}
The argument to characterize $\mathcal{C}_{\max}^{*}=\overline{\text{cone} \lbrace \pi D^{(\mathbf{u})} : \mathbf{u}\in \mathbb{M}^* \rbrace}$  repeats identically. Since $D_0,D_1,D_2$ preserve $\mathcal{K}_{\exp}^*$, $\mathcal{C}_{\max}^{*}\subseteq \mathcal{K}_{\exp}^{*}$. On the other hand, asking that $D_0$ is such that $\pi D_0 \neq 0$ and choosing $\mu_{03}=-1$, $\mu_{01}=e^{-\mu_{02}-1}$ we obtain
that 
\begin{align}
\mathcal{C}_{\max}^* &=\overline{\text{cone} \lbrace \pi D^{(\mathbf{u})} : \mathbf{u}\in \mathbb{M}^* \rbrace}
=\overline{ \text{cone} \lbrace \begin{pmatrix}
\mu_{01}e^x & \mu_{02}+\mu_{03}x & \mu_{03} 
\end{pmatrix} : x \in \mathbb{R} \rbrace}\nonumber\\&=\overline{\text{cone} \lbrace(e^{-x-1},x,-1),x\in\mathbb R\rbrace}=\mathcal{K}_{\exp}^*,
\end{align}
where the last passage is due to the fact that for any $(y_1,y_2,y_3)\in \text{int}\left(\mathcal{K}_{\exp}^*\right)$,  $(-\frac{y_1}{y_3},-\frac{y_2}{y_3},-1)\in \text{int}\left(\mathcal{K}_{\exp}^*\right)$, but $(-\frac{y_1}{y_3},-\frac{y_2}{y_3},-1)$ is also in the convex hull of $\lbrace(e^{-x-1},x,-1),x\in\mathbb R\rbrace$, by the convexity of the function $e^{-x-1}$.
Note that any stable cone $\mathcal{C}$ has to satisfy $\mathcal{C}_{\min} \subseteq \mathcal{C} \subseteq \mathcal{C}_{\max}$. Thus, by the observations above our quasi-realizations has $\mathcal{K}_{\exp}$ as the only closed stable cone. 
Since $\mathcal{C}_{\min}$ and $\mathcal{C}_{\max}^*$ both span the full three-dimensional space, the quasi-realizations are also regular~\cite{Vidyasagar2014,monras2016}.
Moreover, since $\mathcal{K}_{\exp}$ is not semi-algebraic, by the conditions in~\cite{monras2016} the quasi-realization does not admit a completely positive realization.

Now considering the arguments above, we can give a specific example with $a= e$, $b=\frac{1}2$ and 
\begin{align}
m_0^T= \begin{pmatrix}
1 & 1 & 0
\end{pmatrix}\quad
\mu_0 =
\begin{pmatrix}
 1 & -1 & -1
\end{pmatrix},
\end{align}
satisfying the conditions that $\pi m_0>0$ and $\mu_0\tau>0$. 
As a check of consistency, notice that since $\C_{\min}$ and $\C_{\max}$ span $\mathbb R^3$ and $\pi D^{(\mathbf{u})}\tau\geq 0$ for every $\mathbf u\in \mathbb{M}^*$, there must exist a word $\mathbf u^*$ such that $\tau D^{(\mathbf{u}^*)}\pi> 0$, which implies that $\pi\tau>0$, otherwise the probabilities would be all zero. In practice, this is shown already by $\tau D_0 \pi>0$.
We can compute the following fixed points (up to normalization)
\begin{align}
    \tau = \begin{pmatrix}
     17.855... & 5.959... & 1
    \end{pmatrix}^{\!T},\quad
    \pi = \begin{pmatrix}
     2.996...& -1.167... & -1
    \end{pmatrix},
\end{align}
and numerically check that $D_0\tau\neq 0$,$\pi D_0\neq 0$.   

We can then check that $\tau \in \text{int}\left(\mathcal{K}_{\exp}\right)$ and $\pi\in \text{int}\left( \mathcal{K}_{\exp}^*\right)$ explicitly using the expressions \eqref{eq:exponentialcone:app} and \eqref{eq:dualexponentialcone}, which must be true in general because our quasi-realization has minimum dimension ($3$) among all for the generated process.

\subsection{Proof of Theorem~\ref{theopow}: Power cone}
Using the same techniques as before we can give a quasi-realization that does not admit a quantum realization using a reset matrix and diagonal invertible matrices. Since the reasoning is very similar to the one of the previous section the argument is streamlined.
We consider the quasi-realization on $\mathcal{V} = \mathbb{R}^3$ with alphabet $\mathbb{M} = \lbrace 0,1,2,3 \rbrace$ and generators
\begin{align}
D_1 &= \nu'\begin{pmatrix}
 a & 0 & 0 \\
 0 & 1 & 0 \\
 0 & 0 & a^{\frac{\alpha}{\alpha-1}}
\end{pmatrix},
\quad
D_2 = \nu' \begin{pmatrix}
 b & 0 & 0 \\
 0 & 1 & 0 \\
 0 & 0 & b^{\frac{\alpha}{\alpha-1}}
\end{pmatrix},\\
D_3&= \nu'\begin{pmatrix}
 1 & 0 & 0 \\
 0 & -1 & 0 \\
 0 & 0 & 1
\end{pmatrix},
\quad
D_0 = \nu' m_0 \mu_0^T
\end{align}
where $a >1  > b  >0,$ $a+b\neq 1$, $a^{\frac{\alpha}{\alpha-1}}+b^{\frac{\alpha}{\alpha-1}} \neq 1$,$a+b\neq a^{\frac{\alpha}{\alpha-1}}+b^{\frac{\alpha}{\alpha-1}}$, $\frac{\log a}{\log b} \in \mathbb{R}\setminus \mathbb{Q}$ (incommensurate), $\alpha\in\mathbb{R}\setminus\mathbb{Q}$, $0<\alpha <1$ and $\nu'$ is such that the maximum absolute value of the eigenvalues of $D_0+D_1+D_2+D_3$ is $1$.

The power cone and its dual are given by (see Section 4 in \cite{chares2009cones})

\begin{align}
\mathcal{K}_{\alpha} = \left\{ (x_1,x_2,x_3)\in\mathbb{R}^3 : x_1\geq 0,\, x_3\geq 0,\, x_1^{\alpha}x_3^{1-\alpha}\geq |x_2| \right\},
\end{align}

\begin{align}
    \left( \mathcal{K}_{\alpha} \right)^* = B_\alpha \cdot \mathcal{K}^n_{\alpha},
\end{align}
where 

\begin{align}
    B_{\alpha} = \begin{pmatrix}
   \alpha & 0 & 0 \\
    0 & 1 & 0 \\
    0 & 0 & 1-\alpha
    \end{pmatrix}.
\end{align}

Observe that choosing $m_0\in\mathcal{K}_{\alpha},\mu_0^T\in \left( \mathcal{K}_{\alpha} \right)^*$, $D_u,u=0,1,2,3$ preserve the power cone acting from the left and preserve its dual acting from the right. 
Therefore we can find stationary states $\pi\in \mathcal{K}_{\alpha}^*$ and $\tau\in \mathcal{K}_{\alpha}$, and $\mathcal{C}_{\min}\subseteq \mathcal{K}_{\alpha}$, $\C_{\max}^*\subseteq \mathcal{K}_{\alpha}^*$.


Now note that
\begin{align}
    D_1^t D_2^s D_3^k \propto \begin{pmatrix}
    x & 0 & 0 \\
    0 & (-1)^k & 0 \\
    0 & 0 & x^{\frac{\alpha}{\alpha - 1}}
    \end{pmatrix},
\end{align}
where $x = a^t b^s$, which is dense in $\mathbb{R}^+$ due to the incommensurability condition and
Kronecker’s Theorem.

Using that $a+b\neq 1$ and $a^{\frac{\alpha}{\alpha-1}}+b^{\frac{\alpha}{\alpha-1}} \neq 1$ we have that if $D_0 \tau = 0$ then

\begin{align}
    \tau \in  L = \text{span} (1,0,0)^T \qquad\text{or}\qquad  \tau \in  L = \text{span} (0,1,0)^T \qquad \text{or} \qquad  \tau \in  L = \text{span} (0,0,1)^T .
\end{align}
We can use the above observation in order to choose $D_0$ such that $D_0 \tau\neq 0$, having a contradiction. In a similar way we can ensure $\pi D_0\neq 0$. Using the reasoning of Section~\ref{theoexp} we can argue that the boundaries of the minimal and maximal stable cones are generated by $D_1^t D_2^s D_3^k \tau$ and we have 

\begin{align}
    \overline{\mathcal{C}_{\min}} =\overline{ \text{cone} \lbrace \begin{pmatrix}
m_{01}x & \pm m_{02} & m_{03}x^{\frac{\alpha}{\alpha -1}} 
\end{pmatrix} : x >0\rbrace}.
\end{align}

Similarly, the boundaries of the maximal stable cone are generated by $\pi D_0 D_1^t D_2^s D_3^k$ and 
\begin{align}
    \mathcal{C}_{\max}^* =\overline{ \text{cone} \lbrace \begin{pmatrix}
\mu_{01}x & \pm \mu_{02} & \mu_{03}x^{\frac{\alpha}{\alpha -1}} 
\end{pmatrix} : x >0 \rbrace} .
\end{align}





Note that
    \begin{align}
 t\geq s^{\frac{\alpha-1}{\alpha}} \iff (t,\pm 1,s)\in \mathcal{K}_{\alpha} \qquad t\geq \alpha \left(\frac{s}{1-\alpha}\right)^{\frac{\alpha-1}{\alpha}} \iff (t,\pm 1,s)\in \mathcal{K}_{\alpha}^*.  
\end{align}

Thus we can set $\overline{\mathcal{C}_{\min}} = \mathcal{K}_\alpha$ and $\mathcal{C}_{\max}^* = \mathcal{K}^*_\alpha$ by choosing $m_{01} = m_{03}^{\frac{\alpha -1}{\alpha}}$, $m_{02} = \mu_{02} = 1$ and $\mu_{01} = \alpha \left( \frac{\mu_{03} }{1-\alpha}\right)^{\frac{\alpha -1}{\alpha}}$.
Using that any stable cone has to satisfy $\mathcal{C}_{\min} \subseteq \mathcal{C} \subseteq \mathcal{C}_{\max}$ we have that our quasi-realization only has $\mathcal{K}_\alpha$ as a stable cone and by the choice of $\alpha$ it is not semi-algebraic, implying that the quasi-realization cannot have a quantum realization.

\end{appendix}

\end{document}